\documentclass[12pt]{article}
\usepackage{a4wide,epsfig,psfrag,amsmath,amssymb,cite,scalefnt}
\usepackage[dvipsnames]{xcolor}
\usepackage{amsmath,comment,braket}
\usepackage{placeins}
\usepackage{breqn}
\usepackage{slashed}
\usepackage{comment}
\usepackage{subcaption}

\graphicspath{ {figs/} }

\allowdisplaybreaks

\begin{document}

\title{\vskip-3cm{\baselineskip14pt
    \begin{flushleft}
      \normalsize P3H-21-041, TTP21-016
    \end{flushleft}} \vskip1.5cm
  A semi-analytic method to compute Feynman integrals
  applied to four-loop corrections to the
  $\overline{\rm MS}$-pole quark mass relation
}

\author{
  Matteo Fael$^{a}$,
  Fabian Lange$^{a,b}$,
  Kay Sch\"onwald$^{a}$,
  Matthias Steinhauser$^{a}$
  \\[1mm]
  {\small\it $^a$Institut f{\"u}r Theoretische Teilchenphysik}\\
  {\small\it Karlsruhe Institute of Technology (KIT)}\\
  {\small\it Wolfgang-Gaede Stra\ss{}e 1, 76128 Karlsruhe, Germany}
  \\[1mm]
  {\small\it $^b$Institut f{\"u}r Astroteilchenphysik}\\
  {\small\it Karlsruhe Institute of Technology (KIT)} \\
  {\small\it Hermann-von-Helmholtz-Platz 1, 76344 Eggenstein-Leopoldshafen, Germany}
}

\date{}

\maketitle

\thispagestyle{empty}

\begin{abstract}

  We describe a method to numerically compute multi-loop integrals, depending
  on one dimensionless parameter $x$ and the dimension $d$, in the whole
  kinematic range of $x$.  The method is based on differential equations,
  which, however, do not require any special form, and series expansions
  around singular and regular points.  This method provides results well
  suited for fast numerical evaluation and sufficiently precise for
  phenomenological applications.  We apply the approach to four-loop on-shell
  integrals and compute the coefficient function of eight colour structures in
  the relation between the mass of a heavy quark defined in the
  $\overline{\rm MS}$ and the on-shell scheme allowing for a second non-zero
  quark mass.  We also obtain analytic results for these eight coefficient
  functions in terms of harmonic polylogarithms and iterated integrals.  This
  allows for a validation of the numerical accuracy.

\end{abstract}


\thispagestyle{empty}

\newpage


\section{Introduction}

The techniques used for the computation of single-scale integrals are quite
advanced. It has become standard to compute massless propagator and massive
vacuum integrals up to four loops (see, e.g.,
Refs.~\cite{Chetyrkin:2015mxa,Baikov:2015tea}). Recently even the master
integrals for massless five-loop propagator integrals have been
computed~\cite{Georgoudis:2021onj}.  On-shell integrals up to four-loop order
have been considered in
Refs.~\cite{Marquard:2015qpa,Marquard:2016dcn,Laporta:2020fog}, where, however,
most integrals are only available in numerical form.

In this paper we discuss an approach to compute multi-loop integrals which
involve two dimensionful scales, $m_1$ and $m_2$, and thus depend on the
dimensionless quantity $x=m_2/m_1$.  Among other things (which are described
below) it requires single-scale integrals as input and thus a routine
application at four-loop order is possible.  Our method allows
to obtain numerical results with the help of differential equations in
the whole kinematic region of $x$. Details are presented in Section~\ref{sec::method}.

There are a number of algorithms in the literature which can be used to
obtain analytic and/or numeric results of multi-loop
integrals. Some of them make heavy use of integration-by-parts
relations and many approaches exploit the power of differential 
or difference equations~\cite{Kotikov:1990kg,Gehrmann:1999as,Caffo:1998du,Henn:2013pwa}.
For example, in Ref.~\cite{Laporta:2001dd} difference equations
are constructed by raising one of the denominators
of a given single-scale master integral to an arbitrary power $x$.
The solution of the difference equations leads to high-precision
numerical results of the original single-scale integral.

The main idea of Ref.~\cite{Blumlein:2017dxp} is to construct difference
equations for the coefficients of the Laurent expansion in $x$
of the master
integrals by plugging in a suitable ansatz into the system of differential
equations.  Once the difference equations are established, a large number of
expansion terms can usually be obtained rather efficiently so that
one can try to find a closed form solution using other methods (see
e.g.\ Ref.~\cite{Blumlein:2009ta}).  However, the method presented in
Ref.~\cite{Blumlein:2017dxp} cannot be applied for points where the master
integrals obey power-log expansions.

An interesting approach to obtain numerical results of loop integrals
has been presented in Ref.~\cite{Liu:2017jxz} where an imaginary mass is added
to all propagators. The differential equations with respect to this mass
are solved numerically using the infinite-mass limit as boundary.

In the approach of Ref.~\cite{Boughezal:2007ny}, the differential
equations are solved numerically after obtaining boundary conditions by
performing expansions in a suitable limit. Afterwards, the numerical
results can be used to determine the coefficients of expansions around
arbitrary points~\cite{Czakon:2020vql}.

Our approach is close to the method presented in Ref.~\cite{Lee:2017qql}.
This approach uses expansions around singular
points together with the system of differential equations to transfer the
information of the integrals from a starting point, $x_0$, where boundary
conditions are available, to another point, $x_1$.  Note, however, that the
program which comes together with Ref.~\cite{Lee:2017qql}, {\tt DESS.m},
requires the system of differential equations in the so-called ``normalized
global Fuchsian form'', which is not required in our algorithm. Furthermore,
the authors of Ref.~\cite{Lee:2017qql} aim for high-precision results with 
several hundred significant digits, necessary to reconstruct
analytic expressions with the help of the PSLQ algorithm~\cite{PSLQ}. On the
other hand, the goal of our method is the construction of an approximation for
all values of $x$.

Another similar approach is discussed in
Refs.~\cite{Francesco:2019yqt,Hidding:2020ytt} and implemented in the code
{\tt DiffExp}~\cite{Hidding:2020ytt}.  It is more
general than our approach in the sense that the differential equations are solved without the
need of an appropriate ansatz in a respective kinematic point. It aims at
integrating multi-scale integrals along line segments.  Our approach is
taylored to problems which depend on only one dimensionless parameter with
known analytic properties and is optimized for multi-loop problems with
large coupled systems of differential equations in mind.

As an application of our method we consider the relation between
a heavy-quark mass defined in the pole (or on-shell) scheme ($m_1^{\rm OS}$)
and the $\overline{\rm MS}$ scheme ($\overline{m}_1$), which is given by
\begin{eqnarray}
   \overline{m}_1  &=& z_m(\mu) m_1^{\rm OS}
   \,,
  \label{eq::OS2MS}
\end{eqnarray}
where $z_m$ is finite and depends on the renormalization scale $\mu$.
For convenience we also introduce the on-shell renormalization constant
via
\begin{eqnarray}
  m_1^0 &=& Z_m^{\rm OS} m_1 ^{\rm OS}
  \,,
\end{eqnarray}
where $m_1^0$ is the bare quark mass. For the pertubative expansion of $z_m$
we write
\begin{eqnarray}
  z_m = 1 + \sum_{n\ge1} z_m^{(n)} \left(\frac{\alpha_s(\mu)}{\pi}\right)^n
  \,,
\end{eqnarray}
with an analogous definition of $Z_m^{\rm OS}$.

\begin{figure}[t]
  \centering
    \begin{tabular}{cccc}
      \includegraphics[width=0.2\textwidth]{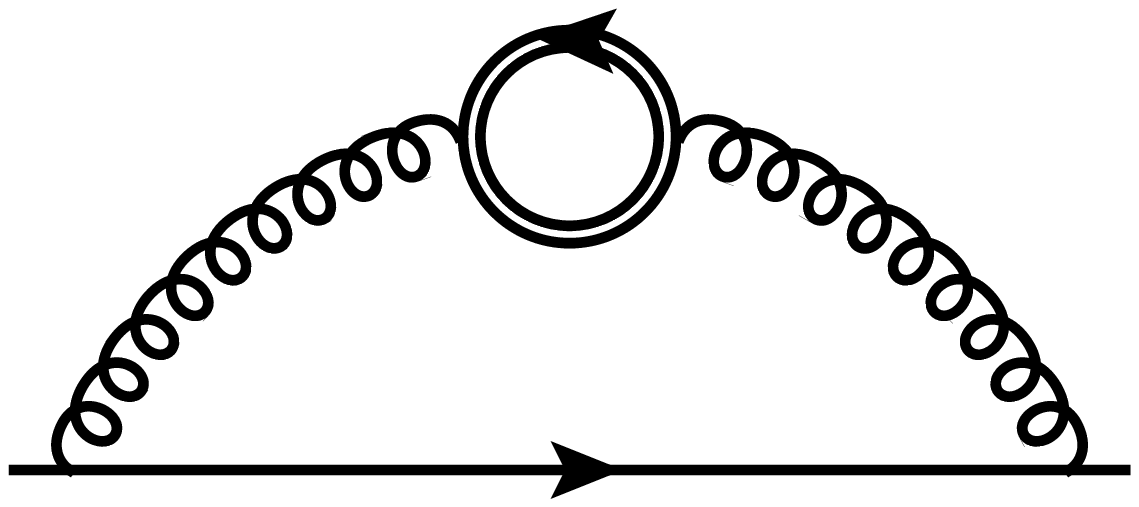} &
      \includegraphics[width=0.2\textwidth]{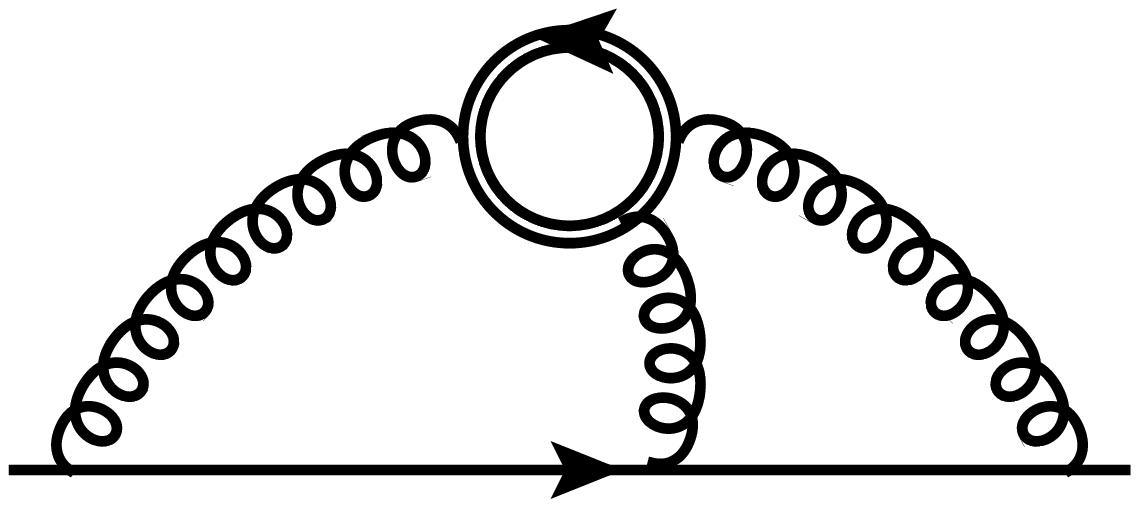} &
      \includegraphics[width=0.2\textwidth]{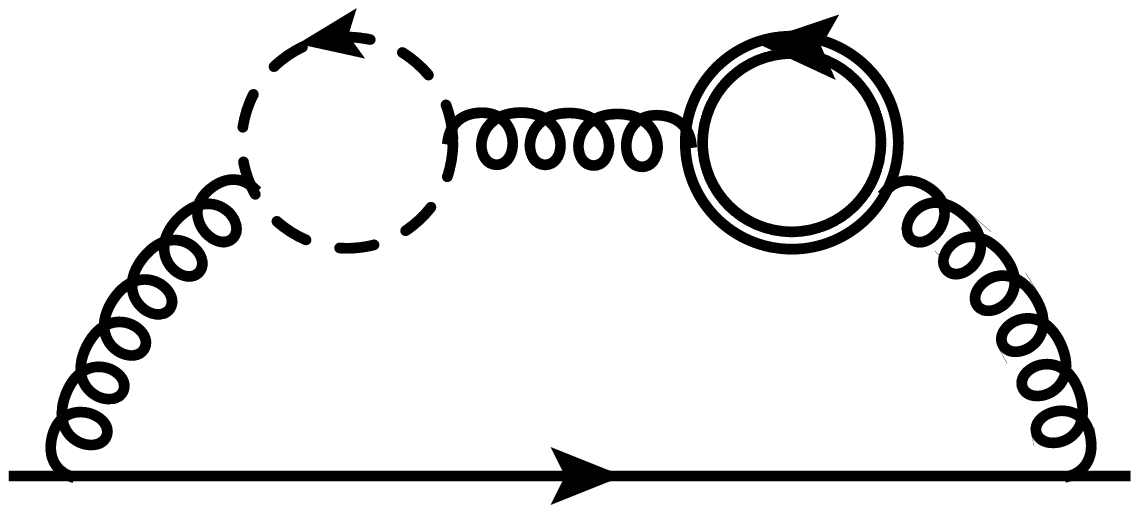} &
      \includegraphics[width=0.2\textwidth]{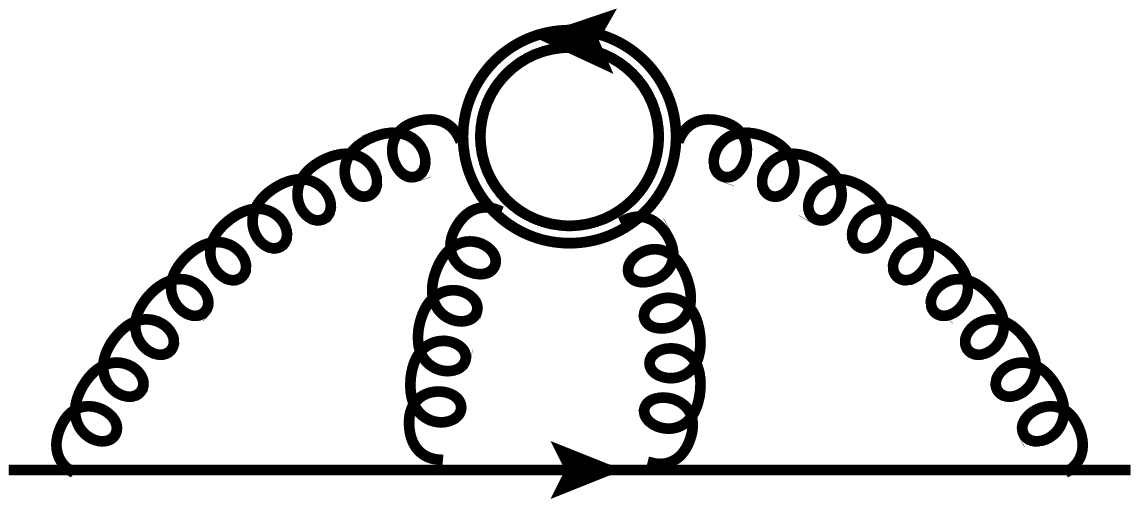} \\
      (a) & (b) & (c) & (d) \\
      \includegraphics[width=0.2\textwidth]{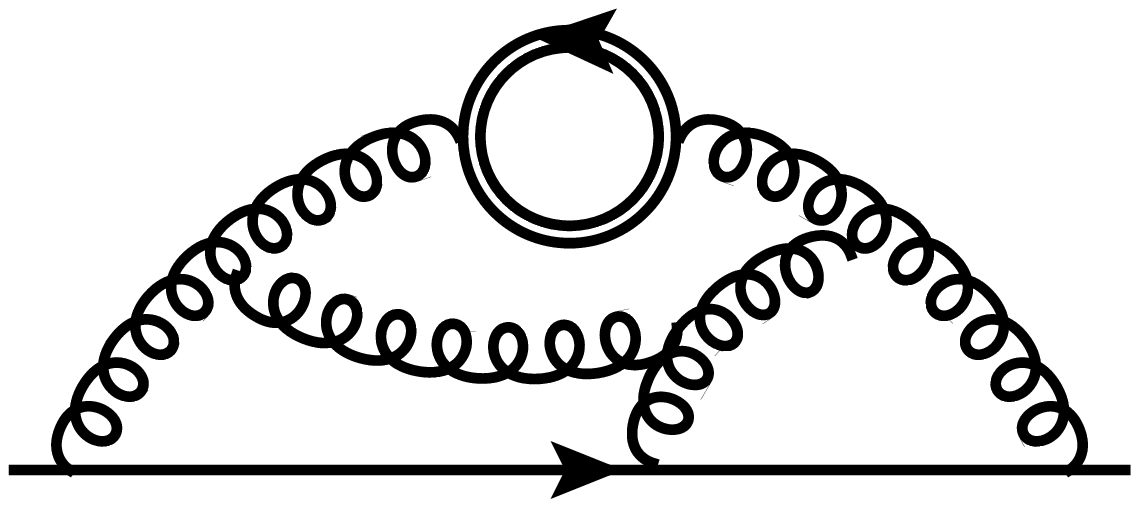} &
      \includegraphics[width=0.2\textwidth]{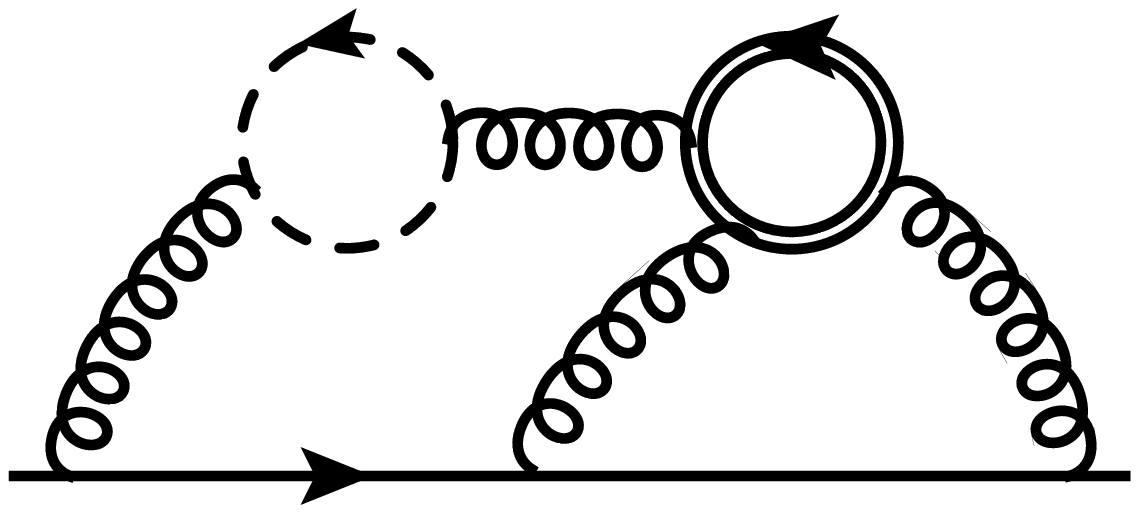} &
      \includegraphics[width=0.2\textwidth]{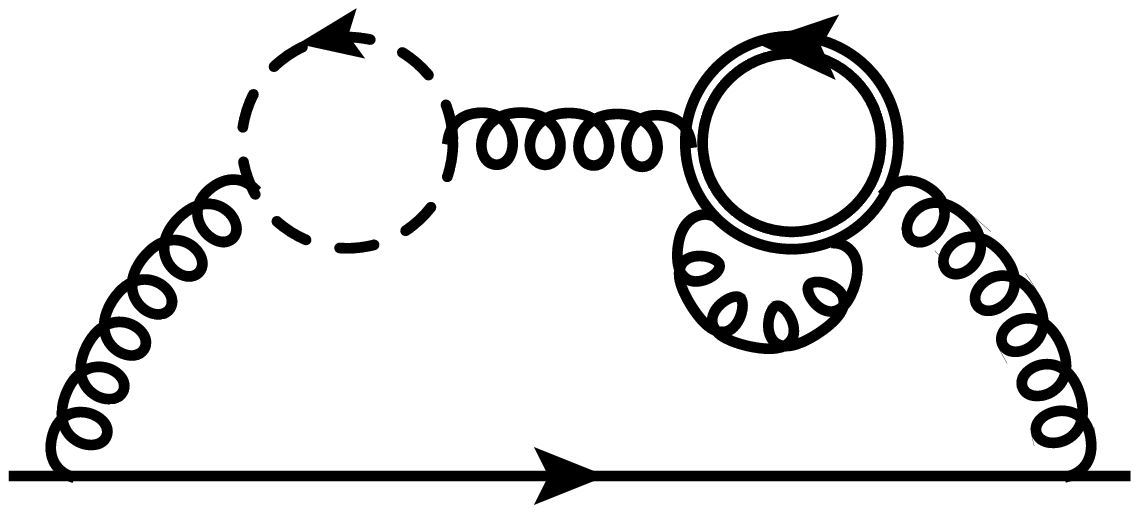} &
      \includegraphics[width=0.2\textwidth]{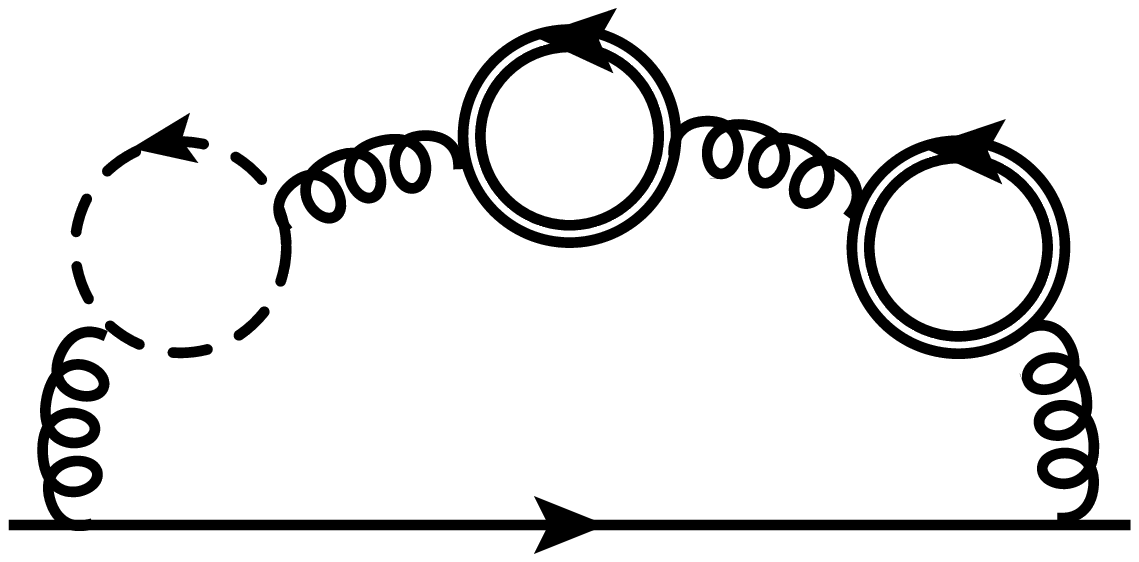} \\
      (e) & (f) & (g) & (h) \\
    \end{tabular}
  \caption{\label{fig::FDs}Sample Feynman diagrams contributing to 
  $Z_m^{\rm OS}$. Straight and curly lines represent
  quarks and gluons, respectively. Dashed and double lines represent massless fermions
  and fermions with mass $m_2$, respectively.}
\end{figure}

Within QCD, analytic results for $z_m$ are available up to three
loops~\cite{Tarrach:1980up,Gray:1990yh,Chetyrkin:1999ys,Chetyrkin:1999qi,Melnikov:2000qh,Melnikov:2000zc,Marquard:2007uj}.
At four-loop order semi-analytic methods were
used~\cite{Marquard:2015qpa,Marquard:2016dcn,Laporta:2020fog}.  Starting from
two loops there are contributions with closed quark loops, which can either be
massless, have the mass of the external quark ($m_1$), or have a different
mass ($m_2$).  Sample Feynman diagrams of this type can be found in
Fig.~\ref{fig::FDs}. The case $0\not= m_2\not=m_1$ was considered in
Refs.~\cite{Gray:1990yh,Broadhurst:1991fy} at two-loop 
and in Refs.~\cite{Bekavac:2007tk,Fael:2020bgs} at three-loop order
(see also
Refs.~\cite{Davydychev:1998si,Grozin:2020jvt}).

In this work we concentrate on the four-loop contributions which involve at
least one additional closed quark loop with mass $m_2$.
We introduce the symbol $n_m$ to count the number of such loops.
In analogy $n_h$ counts the closed loops of fermions with mass $m_1$ 
and $n_l$ the massless ones. To demonstrate our method we 
consider the following eight (out of 16 in total) four-loop colour structures
which involve $n_m$:
\begin{eqnarray}
  z_m^{(4)} &=& 
          C_F T_F^3 n_m n_l^2   z_m^{FMLL}
        + C_F T_F^3 n_m n_l n_h z_m^{FMLH}
        + C_F T_F^3 n_m n_h^2   z_m^{FMHH}
          \nonumber\\&&\mbox{}
        + C_F T_F^3 n_m^2 n_l   z_m^{FMML}
        + C_F T_F^3 n_m^2 n_h   z_m^{FMMH}
        + C_F T_F^3 n_m^3       z_m^{FMMM}
          \nonumber\\&&\mbox{}
        + C_F^2 T_F^2 n_m n_l   z_m^{FFML}
        + C_FC_A T_F^2 n_m n_l  z_m^{FAML}
          \nonumber\\&&\mbox{}
        + \mbox{8 further colour structures involving $n_m$}
          \nonumber\\&&\mbox{}
        + \mbox{23 further colour structures without $n_m$}
                        \,,
                        \label{eq::MSOS_coef}
\end{eqnarray}
with $C_F=(N_c^2-1)/(2N_c)$, $C_A=N_c$ and $T_F=1/2$ for an SU$(N_c)$ gauge
group.  Note that in practical applications we have $n_f=n_l+n_m+n_h=n_l+1+1$
active quark flavours. Numerical result for the coefficients
of Eq.~(\ref{eq::MSOS_coef}) are given in Section~\ref{sec::MSOS}.
For $Z_m^{\rm OS}$ we assume the same decomposition as in Eq.~(\ref{eq::MSOS_coef}).

For the class of integrals considered in this paper we can obtain analytic
results expressed in terms of Goncharov
polylogarithms~\cite{Goncharov:1998kja}. We describe our calculation in
Section~\ref{sec::ana}.


\section{\label{sec::method}Method}

In this section we describe a method to obtain numerical results
of Feynman integrals.  To be concrete we consider a set of master integrals
which depend on the variable $x$ (usually a ratio of two kinematic
invariants) and the dimension $d=4-2\epsilon$.  Let us assume that we are
interested in the results for the integrals for $0\le x < \infty$ and that
an analytic calculation of the master integrals
for $x\gg1$ is possible.

In such cases one often proceeds as follows: One establishes the
differential equations with respect to the variable $x$ for the master
integrals~\cite{Kotikov:1990kg,Gehrmann:1999as}. If they are
sufficiently simple, a direct integration is possible and with the help
of the boundary conditions for large $x$ an exact solution can be
constructed.  Often it is helpful to transform the differential
equations into $\epsilon$- (or canonical)
form~\cite{Henn:2013pwa,Lee:2014ioa} or apply the methods developed in
Refs.~\cite{Ablinger:2015tua,Blumlein:2017dxp,Ablinger:2018zwz}.
There are a number of non-trivial
examples where these approaches have provided analytic results in
terms of harmonic polylogarithms or even more general functions.
However, there are severe limitations.  For example, it is not always
possible to construct an $\epsilon$-form and thus a simple solution of
the differential equations is harder to obtain.  The method, which is
described in the following, does not have such limitations. In fact,
it is, to a large extend, insensitive to the complexity of the
differential equations since only expansions around certain
kinematical points are considered. In particular, it can be applied in
case the differential equations contain elliptic sub-systems.

To apply our method one has to be able to reduce the integrals to a set
of master integrals and to establish a system of differential equations for
the latter. Furthermore, it must be possible to compute the master integrals
in a given limit of the variable $x$. In this limit it is allowed that the
master integrals obey a power-log expansion.  It is not necessary to bring the
system of differential equations into a particular form (e.g. Fuchsian form)
or require that the $x$- and $\epsilon$-dependence factorizes in the
denominators. Furthermore, it is not necessary that the set of master
integrals is minimal.

Our algorithm consists of the following steps:
\begin{itemize}
\item[1.] Reduce all contributing Feynman integrals to master integrals.
\item[2.] Establish the system of differential equations for the master integrals.
\item[3.] Compute boundary conditions for the system of differential
  equations, i.e., evaluate the master integrals for $x$ approaching some
  limit. For clarity, let us consider $x\gg1$.

  In our case, since $m_2$ is an internal mass scale, one can apply the hard
  mass procedure~\cite{Smirnov:2012gma}. This leads to vacuum integrals of the
  considered loop order and products of lower-loop integrals.

\item[4.] Expand the differential equations in this limit and insert an ansatz
  for the master integrals. In general the ansatz is a power-log expansion
  with even and odd powers of $x$.  One can use the boundary conditions to fix
  the constants in the leading term(s) of the ansatz. Afterwards the
  expanded differential equations are used to obtain a deep expansion in
  $1/x$.

  In our application only even powers are present in the limit $x\gg1$.
  Typically we compute 50 expansion terms.

\item[5.] Expand the differential equations for $x\approx1$ and insert an ansatz for
  the master integrals in this limit.

  In our application the expansion around $x=1$ is a simple Taylor expansion.

\item[6.]  Choose a value $x_1$ where both the expansions in $1/x$ and
  around $x=1$ converge.  We evaluate the (known) $1/x$-expanded master
  integrals for $x=x_1$ and use the results as boundary conditions for the
  expansions around $x=1$. A typical value for our application is
  $x_1\approx1.5$.

  Proceeding this way one, in principle, ends up with an over-determined
  system of linear equations which in general has no solution due to the
  numerical errors introduced by truncating the expansions.  To
  circumvent this issue we proceed as follows: We start with the simplest
  master integrals and fix the constant for the leading pole.  We then apply
  this relation to the whole expression and proceed with the next term in the
  $\epsilon$-expansion and repeat this for all master integrals.  If
  more than one unknown constant appears, we solve for one of them.  At some
  point we encounter equations which are linearly dependent on equations
  solved before.  In this case we get a numerical value which would be equal
  to $0$ if we had the exact boundary conditions at $x=x_1$.  We
  store such values and use their absolute value to examine the validity of
  our procedure.  One observes that they get smaller and smaller the more
  terms in the expansion are used.  
  With the help of the differential equations we can
  again compute 50 terms in the expansion around $x=1$.

\item[7.] In a next step we repeat the same procedure to perform a matching
  between the expansions around $x=0$ and $x=1$.

  Note that the expansion of the master integrals around $x=0$ contains again
  both monomials in $x$ and $\log(x)$.  In contrast to the $1/x$ expansion
  both even and odd powers in $x$ are present in general.  For our application
  a proper matching point is $x_0=1/2$.
\end{itemize}
It might happen that the differential equations have further
singularities at $x=x_s$ with $x_s\not=0,1$ or $\infty$, even in case the
physical amplitude does not have thresholds for this value of $x$. In that case
we perform a similar matching at the intermediate value $x=x_s$.
For the integral families considered in this paper we did not encounter
such additional singularities.


\section{\label{sec::ana}$\overline{\rm\bf MS}$-OS relation at four loops:
  analytic results}

For the colour factors which we consider (see
Eq.~\ref{eq::MSOS_coef}) it is possible to obtain analytic results in terms
of iterated integrals.  We use the same techniques already discussed for the
calculation at $\mathcal{O}(\alpha_s^3)$ \cite{Fael:2020bgs} and presented in
Ref.~\cite{Ablinger:2018zwz}.  Let us briefly discuss our procedure.

After generating the amplitudes we map each diagram to an integral
family and express it as a linear combination of scalar functions with
14 arguments, the powers of the scalar propagators. 
We use {\tt Kira}~\cite{Maierhoefer:2017hyi,Klappert:2020nbg} with
\texttt{FireFly}~\cite{Klappert:2019emp,Klappert:2020aqs} for the
reduction to 339 master integrals out of 18 contributing integral
families.\footnote{Even though all 18 families are relatively simple, we
  had to adjust the order of the propagators for two of them according to
  the criteria of Ref.~\cite{Maierhofer:2018gpa} to finish the reductions
  in a reasonable amount of time. Reordering the propagators 
  to a good order increased the performance by a factor of a
  few hundred.}
At that point we
could attempt for a symmetrization of the master integrals over all
different integral families. However, we prefer to consider each
integral family separately.  Of course, there are several master
integrals which are present in different families. We use the
comparison as a cross check.

To obtain the differential equations for this set of master integrals 
we use {\tt LiteRed}~\cite{Lee:2013mka} for the differentiation with 
respect to $x$ and reduce the result again with the help of 
{\tt Kira}. To obtain a closed set 
of differential equations we have to consider a larger set of
integrals than the one present in the amplitude. In total we compute 
the analytic solution of 520 master integrals.
Since our boundary constants are computed in the limit $x \to \infty$
we cast the differential system in the form 
\begin{align*}
  \frac{{\rm d} \vec{M}(z,\epsilon)}{{\rm d}z} 
  &= \mathcal{A}(z,\epsilon) \cdot \vec{M}(z,\epsilon),
\end{align*}
with $z=1/x$ and $\vec{M}(z,\epsilon)$ the vector of master integrals.  
Working with the variable $z$ instead of $x$ allows to fix the boundary at
$z \to 0$ and thus no analytic continuation has to be performed when solving
the differential equations.  In the end, however, an analytic continuation
to the region $z>1$ (i.e. $x<1$) is necessary.

The vector $\vec{M}(z,\epsilon)$ can be chosen in such a way that the matrix
$\mathcal{A}(z,\epsilon)$ is in lower block-triangular form, i.e. the diagonal
elements are square matrices with possible non-vanishing entries to the left.
The square matrices represent coupled systems of master integrals.  We find at
most $5 \times 5$ systems in our calculation.  To solve the coupled systems of
differential equations we utilize \texttt{OreSys}~\cite{ORESYS}, which is
based on \texttt{Sigma}~\cite{Schneider:2007}, to decouple the systems of
equations and obtain a higher-order differential equation for one of the
master integrals in the respective system. Furthermore, we obtain rules which
allow to construct the other master integrals of the coupled system from its
solution.  The higher order differential equation is solved with the help of
\texttt{HarmonicSums}~\cite{HarmonicSums}.  Internally the solver factorizes
the differential equation and, if successful, finds the solution in terms of
iterated integrals without the need of specifying an alphabet.

The boundary constants are fixed using the expansions in the 
limit $x \gg 1$, i.e.\ the limit in which $m_2$ is much 
larger than $m_1$.
For the computation of the boundary conditions for $m_2\gg m_1$ we use the
program ${\tt exp}$~\cite{Seidensticker:1999bb,Harlander:1997zb}
which generates for each master integral all relevant sub- and co-subgraphs
according to the rules of the hard mass procedure~\cite{Smirnov:2012gma}.  
In some cases up to eleven subdiagrams are generated.  
By construction the subdiagrams are one- to four-loop vacuum integrals 
where the relevant scale is given by $m_2$. On the other hand the 
co-subgraphs are propagator-type on-shell integrals up to three loops. 
They only depend on the mass scale $m_1$. All relevant integral families 
are well studied in the literature and the master integrals can be 
found in Refs.~\cite{Laporta:2002pg,Schroder:2005va,Chetyrkin:2006dh,Lee:2010hs,Baikov:2009bg,Heinrich:2009be,Lee:2010ik,Gehrmann:2010ue,Gehrmann:2010tu}.

We compute the expansion for each master integral up to order $1/x^4$.
Note that only a subset of this information is needed in order to
fix the constants in the $x\to \infty$ ansatz. 
Since we decouple the differential equations, we can choose to 
fix the boundary constants by considering the leading term
in the limit $x \to \infty$ of every master integral in the system 
or by considering  higher orders in the expansion for the master 
integral which remains after decoupling. We chose the latter approach.
Still we need at most expansions up to $z^2=1/x^2$. 
All remaining expansion coefficients 
of the master integrals that
we do not need to fix the boundary constants
are used to cross-check the consistency of our results.

The results for the master integrals contributing to the 
amplitude can be written in terms of iterated integrals
\begin{align}
  I \left( \left\{ g(\tau) , \vec{h}(\tau) \right\} ,z\right)
  &=
  \int\limits_{0}^{z} {\rm d}t \, g(t) 
  I \left( \left\{ \vec{h}(\tau) \right\} ,t\right) \,,
\end{align}
with letters drawn from the set 
\begin{align*}
  f_0(\tau) &= \frac{1}{\tau} , &
  f_{1}(\tau) &= \frac{1}{1-\tau} , &
  f_{-1}(\tau) &= \frac{1}{1+\tau} , \\
  f_{w_1}(\tau) &= \sqrt{1-\tau^2} , &
  f_{w_2}(\tau) &= \frac{\sqrt{1-\tau^2}}{\tau} , &
  f_{\{4,0\}}(\tau) &= \frac{1}{1+\tau^2} , &
  f_{\{4,1\}}(\tau) &= \frac{\tau}{1+\tau^2} .
\end{align*}
The first three letters define the harmonic polylogarithms,
the following two were also needed for the calculation at 
$\mathcal{O}(\alpha_s^3)$~\cite{Fael:2020bgs} while the last two cyclotomic 
letters are needed for the $n_m \cdot n_l$
colour factors at $\mathcal{O}(\alpha_s^4)$.

To arrive at a result in the physical region $z = 1/x > 1$ we 
need to analytically continue the iterated integrals.
For the iterated integrals involving the square roots, i.e.
$f_{w_1}$ and $f_{w_2}$, we use differential equations to obtain the 
analytic continuation while the iterated integrals involving 
the other letters can be analytically continued using 
\texttt{HarmonicSums}. Note that the above letters are closed
under the analytic continuation, so we do not need to introduce 
new letters for the physical region. An example how to obtain
the analytic continuation can be found in Ref.~\cite{Fael:2020bgs}.

In the following we present results for $x = 0$ and $x = 1$, which
correspond to particular colour factors in the one-mass limit computed in
Refs.~\cite{Marquard:2016dcn,Laporta:2020fog}. The analytic results in the
variable $x$ are too lengthy to be printed here and can be found in the
ancillary material.  We find
\begin{eqnarray}
  z_m^{FLLL} &=&
  \frac{317 \zeta_3}{432}
  +\frac{71 \pi^4}{4320}
  +\frac{89 \pi^2}{648}
  +\frac{42979}{186624}  
  \,, \nonumber\\
  z_m^{FHHH} &=&
  -\frac{359 \zeta _3}{2160}
  -\frac{13 \pi ^4}{864}
  +\frac{1414991}{933120}
  \,, \nonumber\\
  z_m^{FLLH} &=&
  \frac{5 \zeta _3}{144}
  -\frac{19 \pi ^4}{480}
  +\frac{\pi ^2}{6}
  +\frac{128515}{62208}
  \,, \nonumber\\
  z_m^{FLHH} &=&
  -\frac{623 \zeta _3}{720}
  +\frac{31 \pi ^4}{1440}
  -\frac{2531 \pi ^2}{5400}
  +\frac{1042607}{311040}
  \,, \nonumber\\
  z_m^{FFLL} &=&
  -\frac{88 a_4}{9}
  -\frac{16 a_5}{3}
  +\frac{3 \pi ^2 \zeta _3}{8}
  -\frac{2839 \zeta _3}{576}
  +\frac{305\zeta _5}{48}
  +\frac{2 l_2^5}{45}
  -\frac{11 l_2^4}{27}
  +\frac{4}{27} \pi ^2 l_2^3
  \nonumber \\ &&
  -\frac{22}{27}\pi ^2 l_2^2
  +\frac{31 \pi ^4 l_2}{540}
  +\frac{103 \pi ^2 l_2}{54}
  +\frac{3683 \pi^4}{51840}
  -\frac{5309 \pi ^2}{3456}
  -\frac{2396921}{497664}
  \,, \nonumber\\
  z_m^{FFLH} &=&
  -\frac{8 a_4}{3}
  -\frac{16 a_5}{3}
  -\frac{32 \pi ^2 G}{9}
  +\frac{17 \pi ^2 \zeta_3}{12}
  +\frac{2777 \zeta _3}{288}
  -\frac{13 \zeta _5}{24}
  +\frac{2l_2^5}{45}
  -\frac{l_2^4}{9}
  -\frac{2}{27} \pi ^2 l_2^3
  \nonumber \\ &&
  +\frac{43}{27} \pi ^2 l_2^2
  -\frac{49\pi ^4 l_2}{540}
  -\frac{40 \pi ^2 l_2}{9}
  -\frac{2977 \pi ^4}{25920}
  +\frac{185963 \pi^2}{31104}
  -\frac{3252785}{248832}
  \,, \nonumber\\
  z_m^{FALL} &=&
  \frac{44 a_4}{9}
  +\frac{8 a_5}{3}
  -\frac{13 \pi ^2 \zeta _3}{48}
  -\frac{3245 \zeta _3}{576}
  -\frac{41\zeta _5}{24}
  -\frac{l_2^5}{45}
  +\frac{11 l_2^4}{54}
  -\frac{2}{27} \pi ^2 l_2^3
  +\frac{11}{27} \pi^2 l_2^2
  \nonumber \\ &&
  -\frac{31 \pi ^4 l_2}{1080}
  -\frac{103 \pi ^2 l_2}{108}
  -\frac{4723 \pi^4}{51840}
  -\frac{527 \pi ^2}{384}
  -\frac{2708353}{497664}
  \,, \nonumber\\
  z_m^{FALH} &=&
  -2 \pi ^2 a_4
  +\frac{100 a_4}{3}
  +\frac{8 a_5}{3}
  +\frac{16 \pi ^2 G}{9}
  -\frac{11 \zeta_3^2}{16}
  -\frac{47 \pi ^2 \zeta _3}{24}
  +\frac{4777 \zeta _3}{288}
  +\frac{61 \zeta_5}{12}
  \nonumber \\ &&
  -\frac{7}{4} \pi ^2 \zeta _3 l_2
  -\frac{l_2^5}{45}-\frac{1}{12} \pi ^2 l_2^4
  +\frac{25l_2^4}{18}+\frac{1}{27} \pi ^2 l_2^3
  +\frac{1}{12} \pi ^4 l_2^2
  +\frac{245}{54} \pi ^2l_2^2
  +\frac{49 \pi ^4 l_2}{1080}
  \nonumber \\ &&
  -\frac{535 \pi ^2 l_2}{54}
  +\frac{89 \pi^6}{3780}
  +\frac{5633 \pi ^4}{25920}
  +\frac{15649 \pi ^2}{15552}
  -\frac{6250177}{248832}
  \,,
  \label{eq::zm01}
\end{eqnarray}
with $l_2=\log(2)$, $a_i=\text{Li}_i(1/2)$, 
Riemann's $\zeta$-function $\zeta_i=\sum\limits_{j=1}^\infty 1/j^i$, 
and Catalan's constant defined by $G=\sum\limits_{j=0}^\infty
(-1)^j/(2j+1)^2$. In Eq.~(\ref{eq::zm01}) the renormalization scale 
$\mu=m_1^{\rm OS}$ has been chosen.
The values for $z_m^{FLLL}$, $z_m^{FLLH}$, $z_m^{FFLL}$ and $z_m^{FALL}$
agree with the previously known analytic expressions.
The other analytical results are new and agree with the numerical
values obtained in Ref.~\cite{Marquard:2016dcn} which shows that the
uncertainty estimate made there is correct.


\section{\label{sec::MSOS}$\overline{\rm\bf MS}$-OS relation at four loops:
  numerical results}

\begin{figure}[t]
  \begin{center}
    \includegraphics[width=0.5\textwidth]{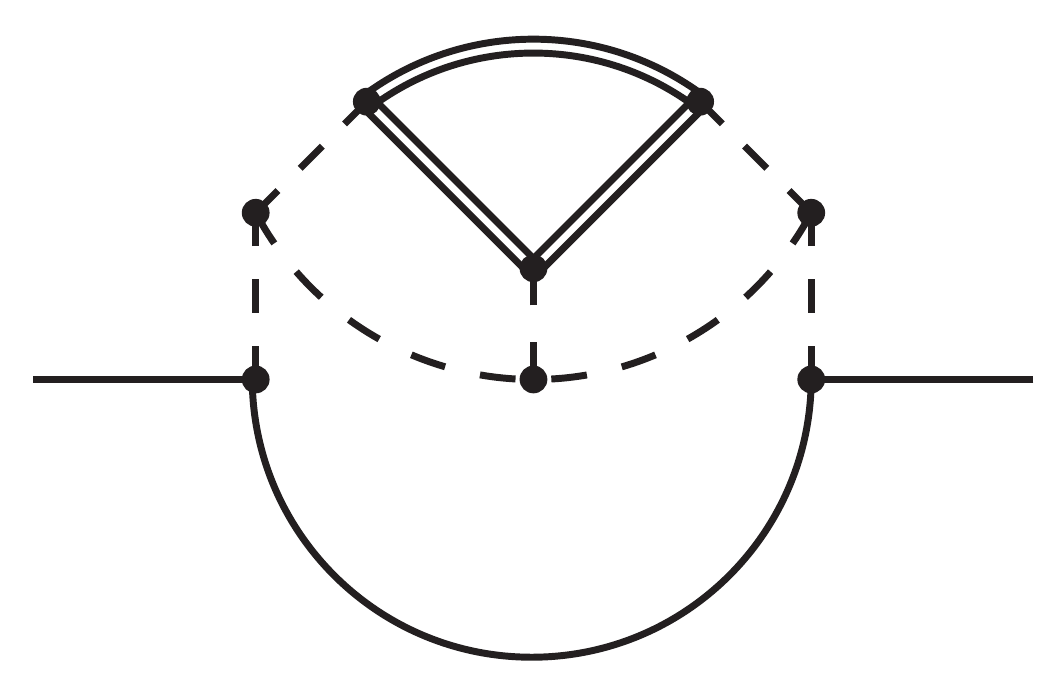}
  \end{center}
  \caption{\label{fig::d4L456} Integral family
    {\tt d4L456}. Dashed, solid and double lines
    represent scalar propagators of mass 0, $m_1$ and $m_2$.}
\end{figure}

In the following we apply the algorithm described in Section~\ref{sec::method}
to the on-shell-$\overline{\rm MS}$ relation.  We start from the same systems
of differential equations and boundary conditions already needed for the
analytic calculation discussed in the previous section.  Note that also in
this approach only a subset of the boundary conditions are needed in order to
fix the constants in the $x\to \infty$ ansatz. In fact, in general we need at
most expansions up to $z^2=1/x^2$. Furthermore, to fix the boundary conditions
of the differential equations not all master integrals are needed. All
remaining coefficients are used to cross-check the consistency of our results.
For example for family {\tt d4L456} (see Fig.~\ref{fig::d4L456}),
the family which introduces the cyclotomic letters in the analytical result,
we find 33 master integrals.  Only the boundary constants of 18 master
integrals, expanded at most up to the constant contribution in the large-$m_2$
limit, are needed.

For some families we observe spurious poles from the reduction up to
$1/\epsilon^4$, which requires an expansion of the boundary integrals up
to order $\epsilon^4$.  For most of the three-loop on-shell and four-loop 
tadpole integrals, which appear in the boundary conditions, an
expansion to such high order is not available. We parameterize the unknown
coefficients and check that they drop out in the physical result.
Alternatively, we could have used the algorithm in
Ref.~\cite{Chetyrkin:2006dh} in order to construct an $\epsilon$-finite basis.
However, for the colour structures considered in this paper this was not
necessary.

\begin{figure}[t]
  \begin{center}
    \begin{tabular}{cc}
      \includegraphics[width=0.5\textwidth]{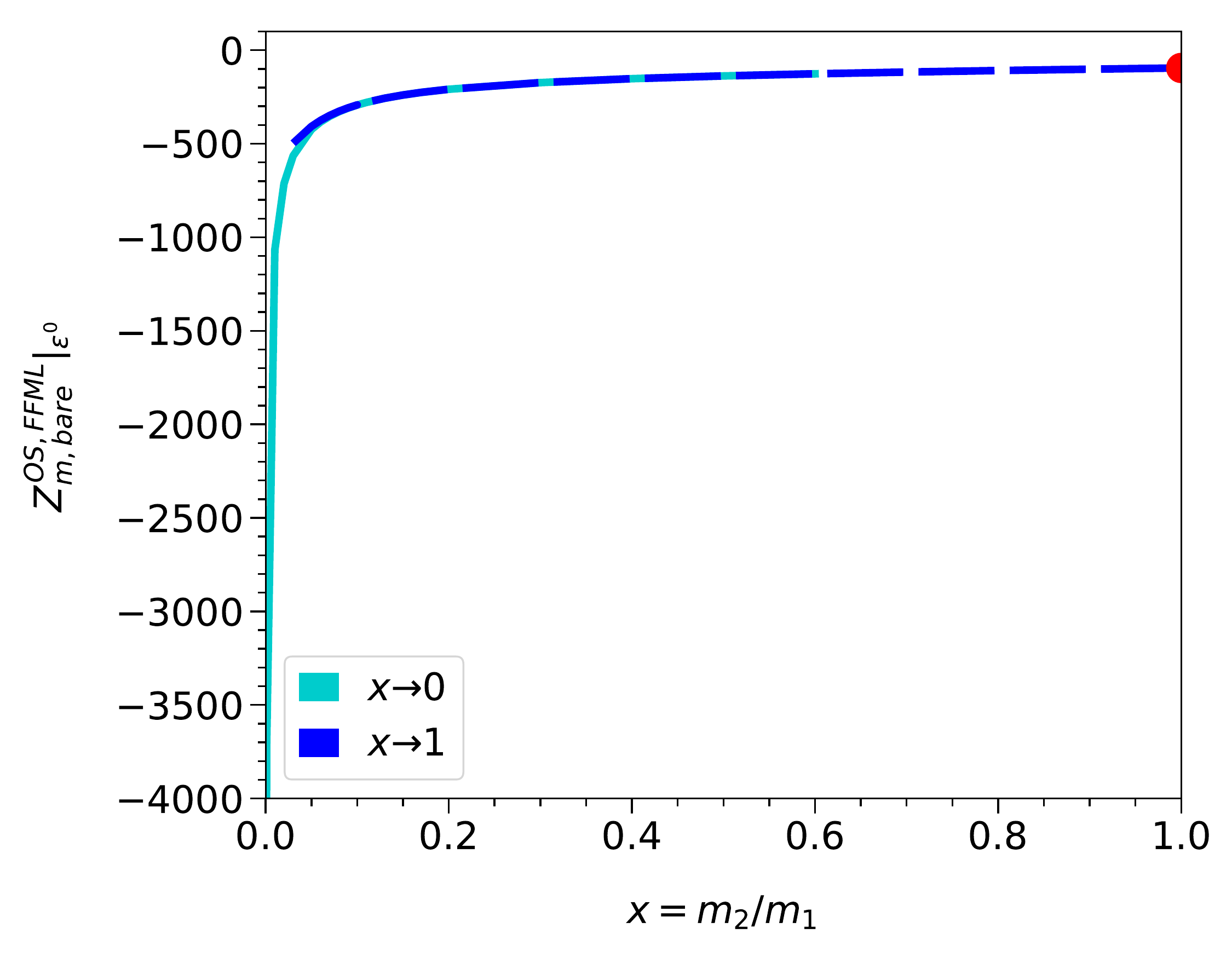} &
      \includegraphics[width=0.5\textwidth]{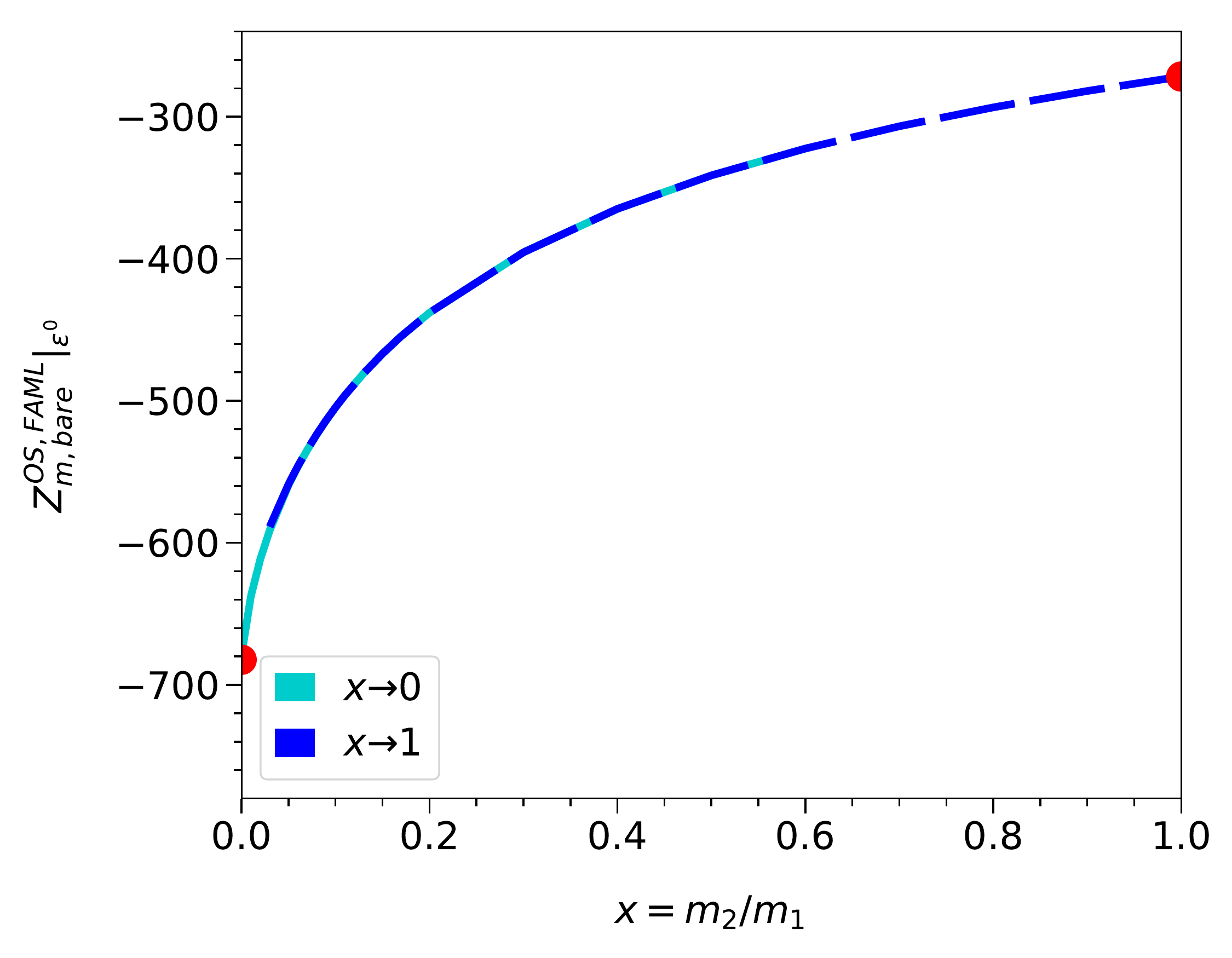} \\
      (a) & (b) 
    \end{tabular}
  \end{center}
  \caption{\label{fig::bare} Bare four-loop results for the colour structures
    $C_F^2n_mn_l$ (a) and $C_FC_An_mn_l$ (b).}
\end{figure}

\begin{figure}[t]
  \begin{center}
    \begin{tabular}{cc}
      \includegraphics[width=0.5\textwidth]{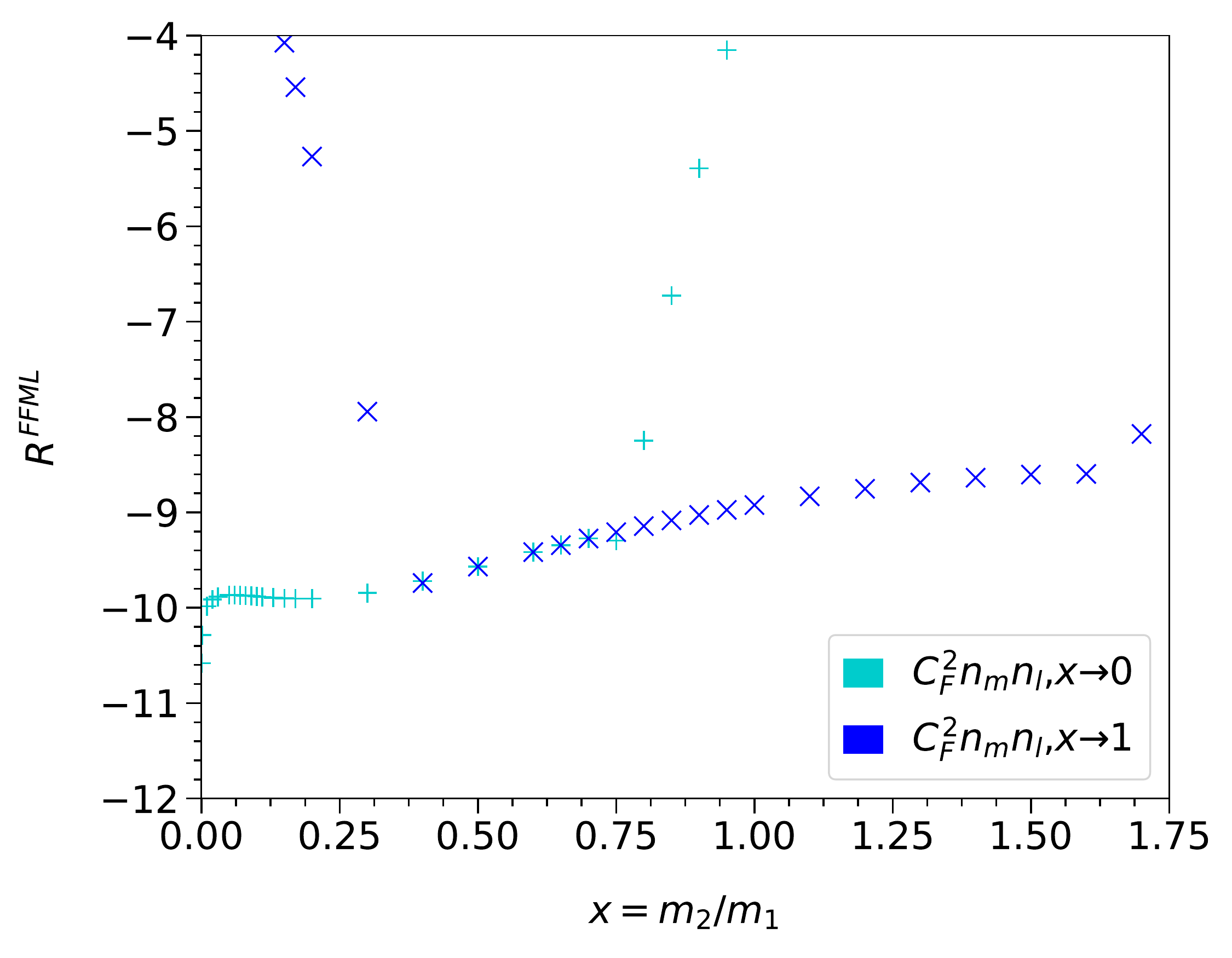} &
      \includegraphics[width=0.5\textwidth]{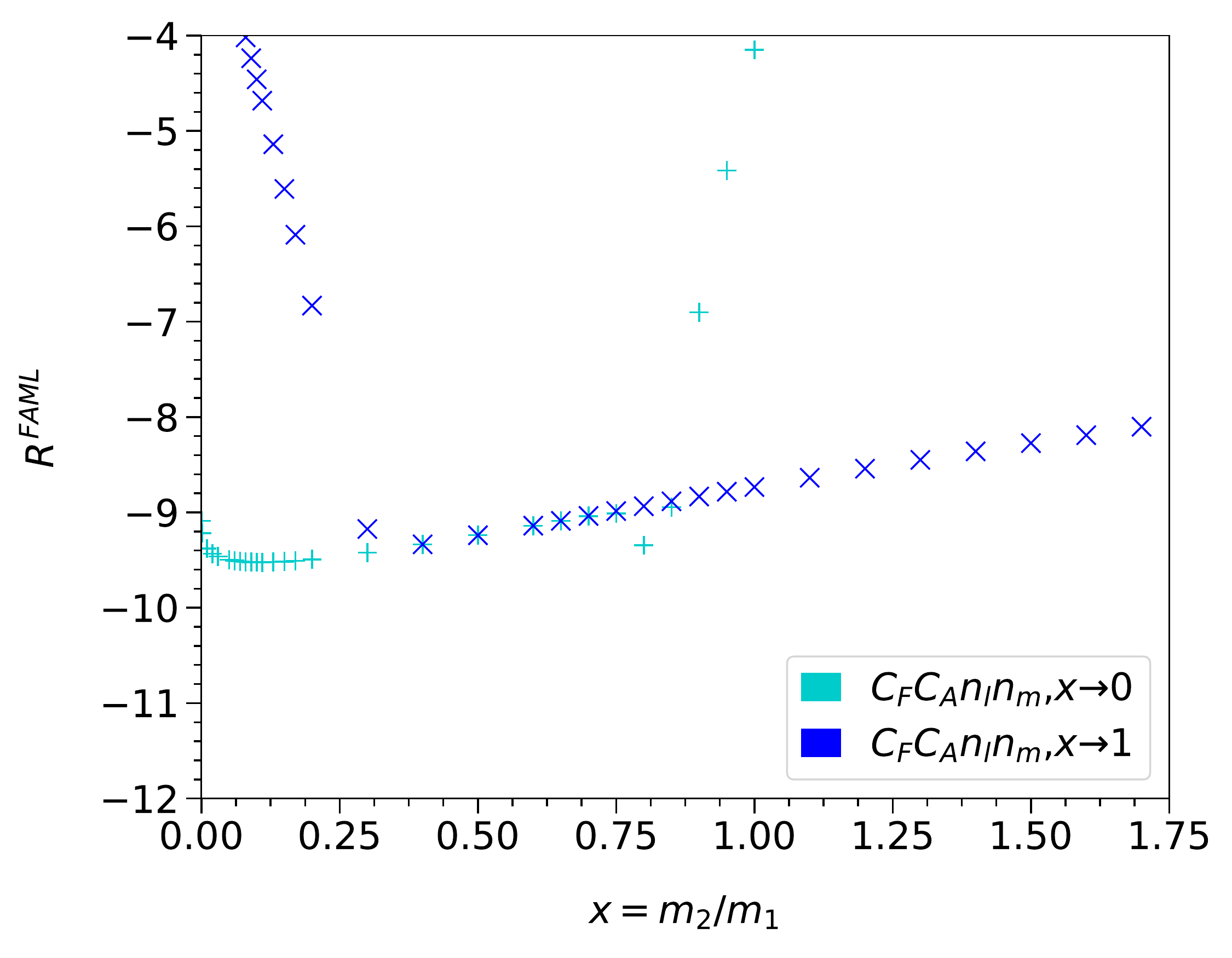} \\
      (a) & (b) 
    \end{tabular}
  \end{center}
  \caption{\label{fig::bare_rel_uncert}Relative uncertainty of bare result
    for the colour structures $C_F^2n_mn_l$ (a) and $C_FC_An_mn_l$ (b).
    See the text for details.}
\end{figure}

We are now in the position to discuss our results. 
For the renormalization scale we again use $\mu=m_1^{\rm OS}$.
We start with the
bare expressions and show in Fig.~\ref{fig::bare} the results of the
$\epsilon^0$ term for the
colour structures $C_F^2n_mn_l$ and $C_FC_An_mn_l$ for $0\le x\le 1$.
The light and dark blue curves show the expansion results around $x=0$
and $x=1$, respectively. The red dots denote the known results for
$m_2=0$ and $m_2=m_1$~\cite{Lee:2013sx,Marquard:2016dcn}.
Note that in the case of $C_F^2n_mn_l$ the limit $x\to0$ does not exist
for the bare expression. In fact, there are logarithmic divergences
which arise from diagrams containing a fermion self energy (see
Fig.~\ref{fig::FDs}(g) for an example). The corresponding analytic
expression is given by
\begin{eqnarray}
    {Z_{m, {\rm bare}}^{{\rm OS},FFML}} &=& 
    -\frac{160}{3 \epsilon ^4}
    + \frac{1}{\epsilon ^3}
    \biggl( 
      48\log(x) 
      - \frac{730}{3} 
    \biggr)
    +\frac{1}{\epsilon ^2} 
    \biggl(
      -128 \log ^2(x)
      +\frac{104}{3}\log (x)
      -\frac{11248}{9}
      \nonumber \\ &&
      -\frac{1384 \pi ^2}{9}
      +\frac{256 \pi ^2 l_2}{3}
      -32 \zeta_3
    \biggr)
    +\frac{1}{\epsilon } 
    \biggl(
       \frac{704}{3} \log ^3(x)
      -\frac{208}{3} \log ^2(x)
      \nonumber \\ &&
      + \biggl[
        \frac{4436}{9}
        + \frac{80 \pi^2}{3}  
      \biggr] \log (x)
      -\frac{103955}{18}
      -\frac{3610 \pi ^2}{3}
      +\frac{7768 \pi ^4}{135}
      -\frac{10240 a_4}{3}
      \nonumber \\ &&
      +\frac{10240 \pi ^2 l_2}{9}
      -\frac{2560}{9} \pi ^2 l_2^2
      -\frac{1280 l_2^4}{9}
      -\frac{35216 \zeta_3}{9}
    \biggr)
    -\frac{992}{3} \log ^4(x)
    +\frac{832 \log ^3(x)}{9}
    \nonumber \\ &&
    - \biggl(
      \frac{11944}{9}
      + 64 \pi^2
    \biggr) \log ^2(x)
    +\biggl(
      - \frac{1606}{27}
      + \frac{104 \pi ^2}{3}
      + 320 \zeta_3
    \biggr) \log (x) 
    -\frac{2440417}{81}
    \nonumber \\ &&
    -\frac{207710 \pi ^2}{27}
    +\frac{63892 \pi ^4}{405}
    -\frac{409600 a_4}{9}
    -\frac{102400 a_5}{3}
    +\frac{69376 \pi ^2 l_2}{9}
    -\frac{256}{9} \pi ^4 l_2
    \nonumber \\ &&
    -\frac{102400}{27} \pi ^2 l_2^2
    +\frac{25600}{27} \pi ^2 l_2^3
    -\frac{51200 l_2^4}{27}
    +\frac{2560 l_2^5}{9}
    -\frac{274048 \zeta_3}{9}
    +\frac{3616 \pi ^2 \zeta_3}{3}
    \nonumber \\ &&
    +\frac{115808 \zeta_5}{3} 
    + {\cal O}(x)\,.
                               \label{eq::ZmOSbare}
\end{eqnarray}
The counterterms cancel all $\log^n(x)$ terms such that for the
renormalized quantities $Z_m^{{\rm OS}, FFML}$ and  $z_{m}^{FFML}$ the limit $x\to 0$ exists.  Note
that our approach reproduces about 10 digits of all 
logarithmically enhanced coefficients and the constant
contribution in Eq.~(\ref{eq::ZmOSbare}).

Fig.~\ref{fig::bare_rel_uncert} shows the relative uncertainty
of our numerical expansion around $x\to0$ and $x\to1$ defined by
\begin{eqnarray}
  R^X &=& \log_{10}\left|
  \frac{Z_{m, {\rm bare}}^{{\rm OS},X}|_{\epsilon^0, \rm exact} 
      - Z_{m, {\rm bare}}^{{\rm OS},X}|_{\epsilon^0, \rm approx.}}
  { Z_{m, {\rm bare}}^{{\rm OS},X}|_{\epsilon^0, \rm exact} }
  \right|
  \,,
\end{eqnarray}
where $X\in\{FFML, FAML\}$.
For both colour structures we observe that the $x\to0$ expansion agrees
with the exact results with 9 to 10 digits up to $x\approx 0.75$.
For larger values of $x$ the quality of the approximation deteriorates quickly.
Similarly, the expansion around $x\to 1$ provides
precise results for $0.25\lesssim x \lesssim 1.7$.

\begin{figure}[t]
  \begin{center}
      \includegraphics[width=0.5\textwidth]{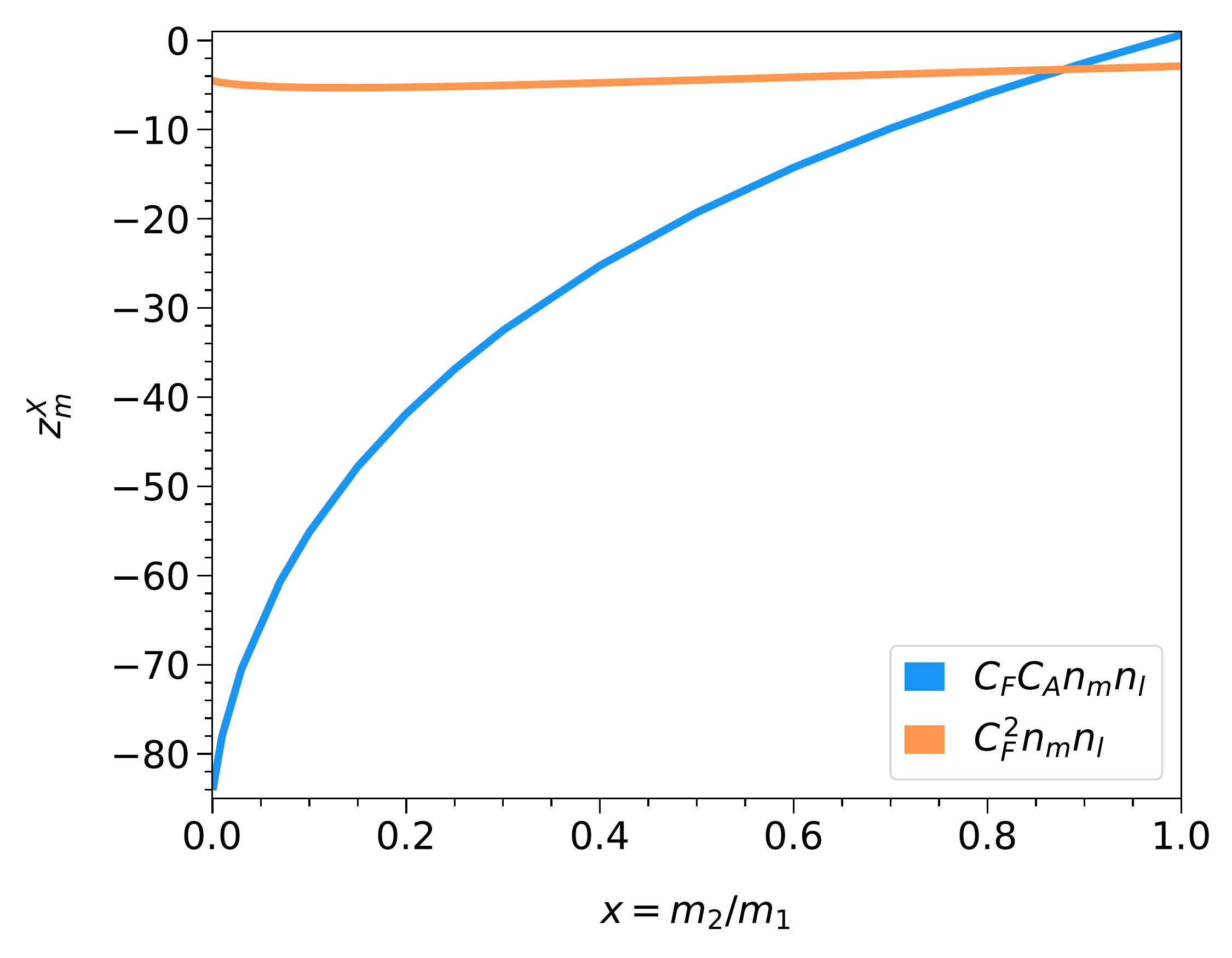}
  \end{center}
  \caption{\label{fig::nmnl_ren} Renormalized results for the
    coefficients $z_{m}^{FFML}$ and $z_{m}^{FAML}$.}
\end{figure}

Fig.~\ref{fig::nmnl_ren} shows the renormalized results for
$z_{m}^{FFML}$ (orange) and $z_{m}^{FAML}$ (blue). The comparison with
Fig.~\ref{fig::bare} shows that there is a substantial cancellation of
more than an order of magnitude between the bare expressions and the
counterterm contributions. Nevertheless, we manage to reproduce at
least five digits of the exact result.

\begin{figure}[t]
  \begin{center}
    \begin{tabular}{cc}
      \includegraphics[width=0.5\textwidth]{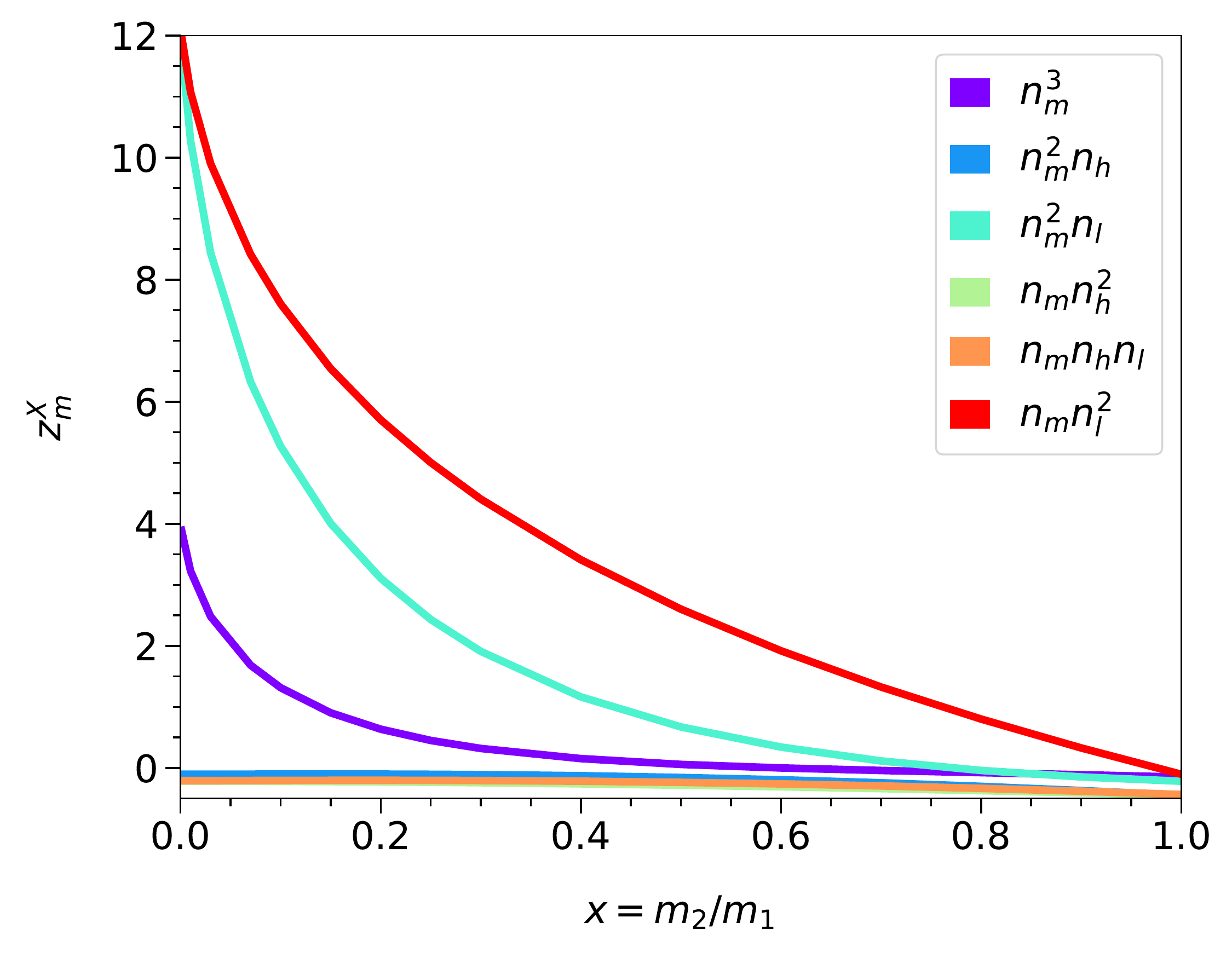}     &
      \includegraphics[width=0.5\textwidth]{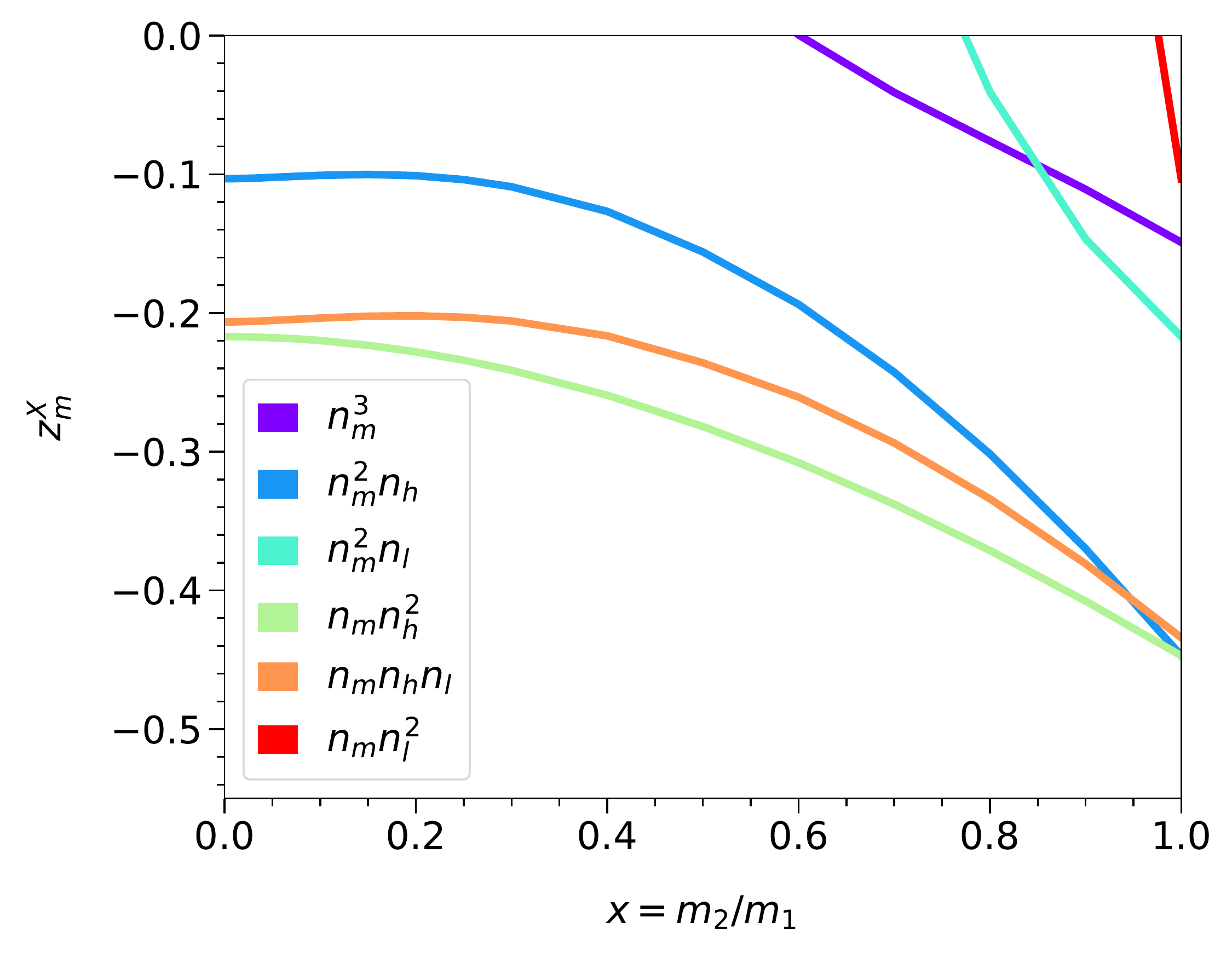} \\
      (a) & (b)
    \end{tabular}
  \end{center}
  \caption{\label{fig::nf3_ren} The coefficients of
    all $n_f^3$ colour structures. (b) is a magnification of (a).}
\end{figure}

The renormalized results for the six $n_f^3$ colour factors are
shown in Fig.~\ref{fig::nf3_ren}. Also here we reproduce the
exact result with a precision between 5 and 10
digits.

The results discussed above are based on expansions around $x=0,1$ and
$\infty$ involving 50 terms. We have repeated the analysis also with fewer
expansion terms and observe a four- to six-significant digit agreement with
the exact expressions in case we use 45 terms for the matching.  One quickly
loses precision in case even less terms are used which is mainly due to the
inability to match the $x\to\infty$ and $x\to1$ expansions properly.
This problem could probably be cured by introducing further matching points.
We have checked that, in case we increase the number of expansion terms
in the matching step from 50 to 60, the agreement with the exact result increases
by about one significant digit.


\section{\label{sec::concl}Conclusions}

In this paper we present a numerical method to compute
multiloop integrals with two mass scales, i.e., one
dimensionless parameter.  The prerequisites necessary to apply our
algorithm are quite simple: It is necessary that the reduction problem
can be solved, that the differential equations can be established and
that boundary conditions can be computed for some limit of
the dimensionless parameter.  We have shown that our approach works
for systems of differential equations which involve a few hundred
master integrals.

As an application we have considered eight colour factors for the
four-loop relation between a heavy-quark mass defined in the
$\overline{\rm MS}$ and on-shell scheme where a second quark mass,
$m_2$, is present in a closed loop.  Analytic boundary conditions are
obtained in the large-$m_2$ limit.

For the considered colour structures we were able to obtain analytic results
which allowed us to quantify the numerical precision to about ten digits in
the whole kinematic range.  For some of the other colour factors we have
observed that not all results can be expressed in terms of iterated integrals
and thus an analytic calculation is much more involved or even impossible with
current techniques. However, our numerical approach can be applied.

We have demonstrated that our approach can reproduce
logarithmic divergences with high accuracy. It could thus also
be applied to compute corrections to the electron contribution
of the anomalous magnetic moment of the muon which 
develops $\log(m_e/m_\mu)$  terms in the limit $m_e\ll m_\mu$
(see, e.g., Refs.~\cite{Kinoshita:2004wi,Kurz:2016bau}).
Furthermore, one can consider non-fermionic on-shell
integrals by introducing an artificial mass in such a way
that analytic boundary conditions can be obtained.
We defer further applications to future work.

Our analytic and most of the numeric results can be found 
in the ancillary file to our paper~\cite{progdata}.



\section*{Acknowledgements}  

This research was supported by the Deutsche Forschungsgemeinschaft (DFG,
German Research Foundation) under grant 396021762 --- TRR 257 ``Particle
Physics Phenomenology after the Higgs Discovery''.
The Feynman diagrams were drawn with the help of Axodraw~\cite{Vermaseren:1994je} and
JaxoDraw~\cite{Binosi:2003yf}.










\begin{thebibliography}{99}

%
%

\bibitem{Chetyrkin:2015mxa}
K.~G.~Chetyrkin, J.~H.~K\"uhn, M.~Steinhauser and C.~Sturm,
Nucl. Part. Phys. Proc. \textbf{261-262} (2015), 19-30
%
[arXiv:1502.00509 [hep-ph]].

\bibitem{Baikov:2015tea}
P.~A.~Baikov, K.~G.~Chetyrkin and J.~H.~K\"uhn,
Nucl. Part. Phys. Proc. \textbf{261-262} (2015), 3-18
%
[arXiv:1501.06739 [hep-ph]].

\bibitem{Georgoudis:2021onj}
A.~Georgoudis, V.~Goncalves, E.~Panzer, R.~Pereira, A.~V.~Smirnov and
V.~A.~Smirnov,
[arXiv:2104.08272 [hep-ph]].

\bibitem{Marquard:2015qpa}
P.~Marquard, A.~V.~Smirnov, V.~A.~Smirnov and M.~Steinhauser,
Phys. Rev. Lett. \textbf{114} (2015), 142002
[arXiv:1502.01030 [hep-ph]].

\bibitem{Marquard:2016dcn}
P.~Marquard, A.~V.~Smirnov, V.~A.~Smirnov, M.~Steinhauser and D.~Wellmann,
Phys. Rev. D \textbf{94} (2016), 074025
[arXiv:1606.06754 [hep-ph]].

\bibitem{Laporta:2020fog}
S.~Laporta,
Phys. Lett. B \textbf{802} (2020), 135264
%
[arXiv:2001.02739 [hep-ph]].

\bibitem{Kotikov:1990kg}
A.~V.~Kotikov,
Phys. Lett. B \textbf{254} (1991), 158-164.

\bibitem{Gehrmann:1999as}
T.~Gehrmann and E.~Remiddi,
Nucl. Phys. B \textbf{580} (2000), 485-518
[arXiv:hep-ph/9912329 [hep-ph]].

\bibitem{Caffo:1998du}
M.~Caffo, H.~Czyz, S.~Laporta and E.~Remiddi,
Nuovo Cim. A \textbf{111} (1998), 365-389
[arXiv:hep-th/9805118 [hep-th]].

\bibitem{Henn:2013pwa}
J.~M.~Henn,
Phys. Rev. Lett. \textbf{110} (2013), 251601
%
[arXiv:1304.1806 [hep-th]].

\bibitem{Laporta:2001dd}
S.~Laporta,
Int. J. Mod. Phys. A \textbf{15} (2000), 5087-5159
%
[arXiv:hep-ph/0102033 [hep-ph]].

\bibitem{Blumlein:2017dxp}
J.~Bl\"umlein and C.~Schneider,
Phys. Lett. B \textbf{771} (2017), 31-36
%
[arXiv:1701.04614 [hep-ph]].

\bibitem{Blumlein:2009ta}
J.~Bl\"umlein,
Comput. Phys. Commun. \textbf{180} (2009), 2218-2249
%
[arXiv:0901.3106 [hep-ph]].

\bibitem{Liu:2017jxz}
X.~Liu, Y.-Q.~Ma and C.-Y.~Wang,
Phys. Lett. B \textbf{779} (2018), 353-357
%
[arXiv:1711.09572 [hep-ph]].

\bibitem{Boughezal:2007ny}
R.~Boughezal, M.~Czakon and T.~Schutzmeier,
JHEP \textbf{09} (2007), 072
[arXiv:0707.3090 [hep-ph]].

\bibitem{Czakon:2020vql}
M.~Czakon and M.~Niggetiedt,
JHEP \textbf{05} (2020), 149
[arXiv:2001.03008 [hep-ph]].

\bibitem{Lee:2017qql}
R.~N.~Lee, A.~V.~Smirnov and V.~A.~Smirnov,
JHEP \textbf{03} (2018), 008
%
[arXiv:1709.07525 [hep-ph]].

\bibitem{PSLQ}
H.~R.~P.~Ferguson and D.~H.~Bailey, RNR Technical Report, RNR-91-032;
H.~R.~P.~Ferguson, D.~H.~Bailey and S.~Arno, NASA Technical Report,
NAS-96-005.

\bibitem{Francesco:2019yqt}
F.~Moriello,
JHEP \textbf{01} (2020), 150
[arXiv:1907.13234 [hep-ph]].

\bibitem{Hidding:2020ytt}
M.~Hidding,
[arXiv:2006.05510 [hep-ph]].

\bibitem{Tarrach:1980up}
R.~Tarrach,
Nucl. Phys. B \textbf{183} (1981), 384-396.

\bibitem{Gray:1990yh}
N.~Gray, D.~J.~Broadhurst, W.~Grafe and K.~Schilcher,
Z. Phys. C \textbf{48} (1990), 673-679.

\bibitem{Chetyrkin:1999ys}
K.~G.~Chetyrkin and M.~Steinhauser,
Phys. Rev. Lett. \textbf{83} (1999), 4001-4004
[arXiv:hep-ph/9907509 [hep-ph]].

\bibitem{Chetyrkin:1999qi}
K.~G.~Chetyrkin and M.~Steinhauser,
Nucl. Phys. B \textbf{573} (2000), 617-651
[arXiv:hep-ph/9911434 [hep-ph]].

\bibitem{Melnikov:2000qh}
K.~Melnikov and T.~van Ritbergen,
Phys. Lett. B \textbf{482} (2000), 99-108
[arXiv:hep-ph/9912391 [hep-ph]].

\bibitem{Melnikov:2000zc}
K.~Melnikov and T.~van Ritbergen,
Nucl. Phys. B \textbf{591} (2000), 515-546
[arXiv:hep-ph/0005131 [hep-ph]].

\bibitem{Marquard:2007uj}
P.~Marquard, L.~Mihaila, J.~H.~Piclum and M.~Steinhauser,
Nucl. Phys. B \textbf{773} (2007), 1-18
[arXiv:hep-ph/0702185 [hep-ph]].

\bibitem{Broadhurst:1991fy}
D.~J.~Broadhurst, N.~Gray and K.~Schilcher,
Z. Phys. C \textbf{52} (1991), 111-122.

\bibitem{Bekavac:2007tk}
S.~Bekavac, A.~Grozin, D.~Seidel and M.~Steinhauser,
JHEP \textbf{10} (2007), 006
%
[arXiv:0708.1729 [hep-ph]].

\bibitem{Fael:2020bgs}
M.~Fael, K.~Sch\"onwald and M.~Steinhauser,
JHEP \textbf{10} (2020), 087
[arXiv:2008.01102 [hep-ph]].

\bibitem{Davydychev:1998si}
A.~I.~Davydychev and A.~G.~Grozin,
Phys. Rev. D \textbf{59} (1999), 054023
[arXiv:hep-ph/9809589 [hep-ph]].

\bibitem{Grozin:2020jvt}
A.~G.~Grozin, P.~Marquard, A.~V.~Smirnov, V.~A.~Smirnov and M.~Steinhauser,
Phys. Rev. D \textbf{102} (2020), 054008
[arXiv:2005.14047 [hep-ph]].

\bibitem{Goncharov:1998kja}
  A.~B.~Goncharov,
  Math.\ Res.\ Lett.\  {\bf 5} (1998) 497-516
  [arXiv:1105.2076 [math.AG]].

\bibitem{Lee:2014ioa}
  R.~N.~Lee,
  JHEP {\bf 04} (2015) 108
%
  [arXiv:1411.0911 [hep-ph]].

\bibitem{Ablinger:2015tua}
J.~Ablinger, A.~Behring, J.~Bl\"umlein, A.~De Freitas, A.~von Manteuffel and C.~Schneider,
Comput. Phys. Commun. \textbf{202} (2016), 33-112
%
[arXiv:1509.08324 [hep-ph]].

\bibitem{Ablinger:2018zwz}
J.~Ablinger, J.~Bl\"umlein, P.~Marquard, N.~Rana and C.~Schneider,
Nucl. Phys. B \textbf{939} (2019), 253-291
[arXiv:1810.12261 [hep-ph]].

\bibitem{Smirnov:2012gma}
V.~A.~Smirnov,
``Analytic tools for Feynman integrals,''
Springer Tracts Mod. Phys. \textbf{250} (2012), 1-296
%

\bibitem{Maierhoefer:2017hyi}
P.~Maierh\"ofer, J.~Usovitsch and P.~Uwer,
Comput. Phys. Commun. \textbf{230} (2018), 99-112
%
[arXiv:1705.05610 [hep-ph]].

\bibitem{Klappert:2020nbg}
J.~Klappert, F.~Lange, P.~Maierh\"ofer and J.~Usovitsch,
Comput. Phys. Commun. \textbf{266} (2021), 108024
[arXiv:2008.06494 [hep-ph]].

\bibitem{Klappert:2019emp}
J.~Klappert and F.~Lange,
Comput. Phys. Commun. \textbf{247} (2020), 106951
%
[arXiv:1904.00009 [cs.SC]].

\bibitem{Klappert:2020aqs}
J.~Klappert, S.~Y.~Klein and F.~Lange,
Comput. Phys. Commun. \textbf{264} (2021), 107968
%
[arXiv:2004.01463 [cs.MS]].

\bibitem{Maierhofer:2018gpa}
P.~Maierh\"ofer and J.~Usovitsch,
[arXiv:1812.01491 [hep-ph]].

\bibitem{Lee:2013mka}
R.~N.~Lee,
J. Phys. Conf. Ser. \textbf{523} (2014), 012059
[arXiv:1310.1145 [hep-ph]].

\bibitem{ORESYS}
S.~Gerhold, {\it Uncoupling systems of linear {O}re operator equations},
Diploma Thesis, RISC, J.~Kepler University, Linz, 2002.
%
%

\bibitem{Schneider:2007}
C.~Schneider, {S\'em.~Lothar. Combin.\/} {\bf 56} (2007) 1, 
 article B56b;
C.~Schneider, in:~{{Computer Algebra in Quantum Field Theory: Integration,
  Summation and Special Functions}\/} Texts and Monographs in Symbolic
  Computation eds. C.~Schneider and J.~Bl\"umlein  (Springer, Wien, 2013) 325
  arXiv:1304.4134 [cs.SC].

\bibitem{HarmonicSums}
J.~A.~M.~Vermaseren,
Int. J. Mod. Phys. A \textbf{14} (1999), 2037-2076
%
[arXiv:hep-ph/9806280 [hep-ph]];
%
%
%
%
J.~Bl\"umlein,
Comput. Phys. Commun. \textbf{180} (2009), 2218-2249
%
[arXiv:0901.3106 [hep-ph]];
  J.~Ablinger,
  Diploma Thesis, J. Kepler University Linz, 2009,
  arXiv:1011.1176 [math-ph];
  J.~Ablinger, J.~Bl\"umlein and C.~Schneider,
  J.\ Math.\ Phys.\  {\bf 52} (2011) 102301
  [arXiv:1105.6063 [math-ph]];
J.~Ablinger, J.~Bl\"umlein and C.~Schneider,
J. Math. Phys. \textbf{54} (2013), 082301
%
[arXiv:1302.0378 [math-ph]];
  J.~Ablinger,
  Ph.D. Thesis, J. Kepler University Linz, 2012,
  arXiv:1305.0687 [math-ph];
J.~Ablinger, J.~Bl\"umlein and C.~Schneider,
J. Phys. Conf. Ser. \textbf{523} (2014), 012060
%
[arXiv:1310.5645 [math-ph]];
J.~Ablinger, J.~Bl\"umlein, C.~G.~Raab and C.~Schneider,
J. Math. Phys. \textbf{55} (2014), 112301
%
[arXiv:1407.1822 [hep-th]];
%
J.~Ablinger,
PoS \textbf{LL2014} (2014), 019
%
[arXiv:1407.6180 [cs.SC]];
%
J.~Ablinger,
[arXiv:1606.02845 [cs.SC]];
%
J.~Ablinger,
PoS \textbf{RADCOR2017} (2017), 069
[arXiv:1801.01039 [cs.SC]];
%
J.~Ablinger,
PoS \textbf{LL2018} (2018), 063;
%
%
J.~Ablinger,
[arXiv:1902.11001 [math.CO]].
%

\bibitem{Seidensticker:1999bb}
  T.~Seidensticker,
  hep-ph/9905298.

\bibitem{Harlander:1997zb}
  R.~Harlander, T.~Seidensticker and M.~Steinhauser,
  Phys.\ Lett.\ B {\bf 426} (1998) 125-132
  [hep-ph/9712228].

\bibitem{Laporta:2002pg}
S.~Laporta,
Phys. Lett. B \textbf{549} (2002), 115-122
[arXiv:hep-ph/0210336 [hep-ph]].

\bibitem{Schroder:2005va}
Y.~Schroder and A.~Vuorinen,
JHEP \textbf{06} (2005), 051
[arXiv:hep-ph/0503209 [hep-ph]].

\bibitem{Chetyrkin:2006dh}
K.~G.~Chetyrkin, M.~Faisst, C.~Sturm and M.~Tentyukov,
Nucl. Phys. B \textbf{742} (2006), 208-229
%
[arXiv:hep-ph/0601165 [hep-ph]].

\bibitem{Lee:2010hs}
R.~N.~Lee and I.~S.~Terekhov,
JHEP \textbf{01} (2011), 068
[arXiv:1010.6117 [hep-ph]].

\bibitem{Baikov:2009bg}
P.~A.~Baikov, K.~G.~Chetyrkin, A.~V.~Smirnov, V.~A.~Smirnov and M.~Steinhauser,
Phys. Rev. Lett. \textbf{102} (2009), 212002
[arXiv:0902.3519 [hep-ph]].

\bibitem{Heinrich:2009be}
G.~Heinrich, T.~Huber, D.~A.~Kosower and V.~A.~Smirnov,
Phys. Lett. B \textbf{678} (2009), 359-366
[arXiv:0902.3512 [hep-ph]].

\bibitem{Lee:2010ik}
R.~N.~Lee and V.~A.~Smirnov,
JHEP \textbf{02} (2011), 102
[arXiv:1010.1334 [hep-ph]].
%

\bibitem{Gehrmann:2010ue}
T.~Gehrmann, E.~W.~N.~Glover, T.~Huber, N.~Ikizlerli and C.~Studerus,
JHEP \textbf{06} (2010), 094
[arXiv:1004.3653 [hep-ph]].

\bibitem{Gehrmann:2010tu}
T.~Gehrmann, E.~W.~N.~Glover, T.~Huber, N.~Ikizlerli and C.~Studerus,
JHEP \textbf{11} (2010), 102
[arXiv:1010.4478 [hep-ph]].

\bibitem{Lee:2013sx}
R.~Lee, P.~Marquard, A.~V.~Smirnov, V.~A.~Smirnov and M.~Steinhauser,
JHEP \textbf{03} (2013), 162
%
[arXiv:1301.6481 [hep-ph]].

\bibitem{Kinoshita:2004wi}
T.~Kinoshita and M.~Nio,
Phys. Rev. D \textbf{70} (2004), 113001
[arXiv:hep-ph/0402206 [hep-ph]].

\bibitem{Kurz:2016bau}
A.~Kurz, T.~Liu, P.~Marquard, A.~V.~Smirnov, V.~A.~Smirnov and M.~Steinhauser,
Phys. Rev. D \textbf{93} (2016), 053017
%
[arXiv:1602.02785 [hep-ph]].

\bibitem{progdata}
\verb|https://www.ttp.kit.edu/preprints/2021/ttp21-016/|.

\bibitem{Vermaseren:1994je}
J.~A.~M.~Vermaseren,
Comput. Phys. Commun. \textbf{83} (1994), 45-58.

\bibitem{Binosi:2003yf}
D.~Binosi and L.~Theu{\ss}l,
Comput. Phys. Commun. \textbf{161} (2004), 76-86
[arXiv:hep-ph/0309015 [hep-ph]].


\end{thebibliography}
\end{document}